\documentclass[12pt]{article}
\usepackage{graphicx}
\usepackage{amssymb}
\usepackage{amsmath}
\usepackage{epsf}
\usepackage[usenames]{color}
\usepackage[parfill]{parskip}
\usepackage[matrix,arrow]{xy}
\usepackage{multirow}
\input{epsf}

\newcommand{\ba}{\begin{eqnarray*}}
\newcommand{\ea}{\end{eqnarray*}}
\newcommand{\ban}{\begin{eqnarray}}
\newcommand{\ean}{\end{eqnarray}}

\newcommand{\IC}{\mathbb{C}}
\newcommand{\IP}{\mathbb{P}}

\newcommand{\IR}{\mathbb{R}}

\newcommand{\cM}{{\cal M}}

\newcommand{\cN}{{\cal N}}
\newcommand{\cP}{{\cal P}}
\newcommand{\cD}{{\cal D}}

\newcommand{\cU}{{\cal U}}

\newcommand{\del}{\partial}

\newcommand{\re}{{\rm Re \,}}
\newcommand{\im}{{\rm Im \,}}

\newcommand{\txi}{{\tilde \xi}}
\newcommand{\tP}{{\tilde \cP}}
\newcommand{\tomega}{{\tilde \omega}}
\newcommand{\tmu}{{\tilde \mu}}
\newcommand{\bzeta}{\bar{\zeta}}

\def \nn{\nonumber}

\makeatletter
\@addtoreset{equation}{section}
\makeatother

\begin{document}
\noindent
\begin{titlepage}

\begin{center}
\hfill ITFA-2007-43\\
\vskip 2cm {\Huge
Nearly K\"ahler Reduction\\ \vskip 0.1cm}
\vskip 1cm {Amir-Kian Kashani-Poor} \\ \vskip 0.5cm
{Institute for Theoretical Physics, University of Amsterdam\\
1018 XE Amsterdam, The Netherlands\\} \vskip0.2cm
\end{center} 
\vskip 1.5cm
\begin{abstract}
We consider compactification of type IIA supergravity on nearly K\"ahler manifolds. These represent a simple class of $SU(3)$ structure manifolds which includes $S^6$ and $\IC \IP^3$. We exhibit for the first time an explicit reduction ansatz in this context, obtaining an $\cN=2$ gauged supergravity in 4d with a single vector and hypermultiplet. We verify that supersymmetric solutions of the 4d theory lift to 10d solutions. Along the way, we discuss questions related to encountering both electric and magnetic charges in the 4d theory.
\end{abstract}
\end{titlepage}
\newpage
\section{Introduction}
$SU(3)$ structure manifolds permit a non-vanishing global spinor, hence are a natural starting point for compactifications of type II supergravities to 4 dimensional theories with fermions. This observation has triggered much work aimed at utilizing the constraints arising from the reduced structure group both in the study of the 10d theory and the resulting 4d theory, including \cite{Gurrieri:2002wz,Gurrieri:2002iw,Behrndt:2004km,Behrndt:2004mj,Lust:2004ig,Grana:2005sn,Grana:2005ny,House:2005yc,Grana:2006hr,Louis:2006wq,Cassani:2007pq,Micu:2007rd}  . In \cite{Gurrieri:2002wz}, a program was initiated to perform a reduction on such manifolds, modelled on Calabi-Yau reductions, to obtain an effective four dimensional action described within the framework of $\cN=2$ gauged supergravity. While yielding tantalizing results, the status of this approach is still unclear with regard to the following main points:
\begin{itemize}
\item The reduction algorithm takes as its starting point a system of forms satisfying a set of differential constraints. These forms have not been characterized intrinsically, nor has an explicit system of such forms (aside from the trivial Calabi-Yau case) been exhibited to date.
\item It has not been demonstrated that solutions to the effective 4d theory lift to 10d solutions.
\item It is not clear to what extent this approach captures all light degrees of freedom. 
\end{itemize}
Note the interrelation between these points: if one demonstrates that the expansion forms chosen indeed yield an orbit on field space on which 10d solutions lie, then these solutions will give rise to extrema of the 4d effective action as well, and hence these extrema will lift.  In addition, one might encounter additional 4d extrema that do not lift. To rule out this possibility requires guaranteeing that all light degrees of freedom are captured by the reduction ansatz. 

In this note, we wish to address the first two of these questions in the context of internal manifolds exhibiting nearly K\"ahler structure. The central simplifying feature of this class of $SU(3)$ structure manifolds is that the invariant 2- and 3-forms $J$ and $\Omega$ are eigenforms of the Laplacian associated to the metric they specify. Hence, the set of forms they must be expanded in is clear.\footnote{Note that in the context of 11d supergravity, a similar simplification arises when reducing on weak $G_2$ holonomy manifolds \cite{House:2004pm}.} We will perform the reduction on a 1 dimensional family of nearly K\"ahler structures, and demonstrate that the supersymmetric solution of the resulting 4d gauged supergravity lifts to 10d. 

Here is a summary of the organization and the results of this paper: we review the general setup of $SU(3)$ flux compactifications of type II supergravity to 4d $\cN=2$ gauged supergravity in section \ref{setup}. In section \ref{nearlyK}, we briefly survey some facts on nearly K\"ahler manifolds from the mathematics literature. We turn to the question of the appropriate choice of expansion forms in section \ref{choice of expansion forms}.  We argue in this section that a metric ansatz parametrizing a family of nearly K\"ahler manifolds in conjunction with the ansatz reviewed in section \ref{setup} for the expansion forms constrains us to a reduction which gives rise to 4d gauged supergravity with a single vector multiplet and only the universal hypermultiplet. Note that a richer 4d theory might well be accessible upon weakening either of these two premises. The 4d theory we obtain involves both electric and magnetic gauging, and we discuss the formulation of $\cN=2$ supergravity permitting this structure in section \ref{em gauging}. We also review quaternionic K\"ahler manifolds and the moment map construction in this section. Up to this point, the discussion takes place purely at the level of actions. In a very nice paper, \cite{Cassani:2007pq} demonstrated recently that $\cN=1$ constraints imposed on 4d field configurations lift 
to the corresponding 10d constraints, as worked out e.g. in \cite{Grana:2006kf} (\cite{Koerber:2007xk} perform a similar analysis from an $\cN=1$ point of view). Based on this work, we recover in section \ref{lifting} the nearly K\"ahler field configurations of type IIA supergravity preserving 4 supercharges discussed in the literature \cite{Behrndt:2004km,Behrndt:2004mj,Lust:2004ig,Grana:2006kf} from our 4d action. As is well-known, it is not guaranteed that supersymmetric field configurations solve the equations of motion. \cite{Lust:2004ig} demonstrates in 10d that, up to a minor restriction, supersymmetric field configurations of type IIA preserving 4 supercharges do have this property. In section \ref{lifting}, we provide the required argument in 4d for our setup, and then proceed to demonstrate explicitly that by imposing $\cN=1$ constraints, we obtain a solution to the 4d equations of motion. In appendix \ref{specialK} and \ref{indnot}, we summarize some facts on special K\"ahler manifolds and our conventions and notation. Appendix \ref{Lichnerowicz} lies somewhat outside the main line of development of this note. In it, we complete the proof sketched in \cite{KashaniPoor:2006si} regarding a property of the variation of harmonic 2-forms on Calabi-Yau manifolds.

What is missing in these considerations is an analysis of to what extent our reduction ansatz is capturing all light degrees of freedom of the system (cf. the discussion in \cite{KashaniPoor:2006si}). The consequence of not considering all light modes would be that some (non-supersymmetric) 4d solutions might not lift to 10d solutions: a field configuration minimizing the 4d action could be destabilized in a direction which is omitted from the reduction ansatz. This issue should be settled by considering the reduction ansatz at the level of the equations of motion, along the lines of \cite{Gauntlett:2007ma}.

\section{The setup} \label{setup}
$SU(3)$ structure can be obtained as the intersection of an almost symplectic ($Sp(6,\IR)$) and an almost complex ($SL(3,\IC)$) structure. These structures can in turn be encoded in a 2-form $J$ and 3-form $\Omega$ respectively. This similarity with Calabi-Yau geometry has inspired reduction ans\"atze in the literature, starting with \cite{Gurrieri:2002wz}, in which $J$ and $\Omega$ are expanded in the same set of internal two and three forms as the RR and NS field strengths,
\ba
J &=& v^i \omega_i \,, \\
\Omega &=& Z^A \alpha_A - G_A \beta^A  \,.
\ea
$J$ and $\Omega$ are not closed in general, and in fact, their failure to be closed is parametrized by the 5 torsion classes which specify the $SU(3)$ structure \cite{Chiossi:2002tw},
\ba
dJ &=& -\frac{3}{2}\, {\rm Im}(W_1 \bar{\Omega}) + W_4 \wedge J + W_3 \, ,   \\
d\Omega &=& W_1 J^2 + W_2 \wedge J + \bar{W_5} \wedge \Omega\ \, .   
\ea
Hence, the expansion forms cannot be harmonic forms as in conventional Calabi-Yau reductions. Instead, they were proposed to obey the following differential system \cite{Gurrieri:2002wz,Gurrieri:2002iw,DAuria:2004tr},
\ban
d^\dagger \omega_i &=& 0 \nn \\
d\omega_i &=& m_i{}^A \alpha_A + e_{i A} \beta^A \nn \\
d \alpha_A = e_{iA} \tomega^i &;& d\beta^A = -m_i{}^A \tomega^i \nn \\
d\tomega^i &=& 0  \,.  \label{diffsys}
\ean
Starting with \cite{Gurrieri:2002wz}, the reduction algorithm based on such a system of forms was demonstrated to give rise to 4d $\cN=2$ gauged supergravity, with the integers $e_{i A}$ and $m_i{}^A$ mapping to charges of the 4d matter fields. Much earlier \cite{Polchinski:1995sm}, a reduction algorithm involving undeformed expansion forms in the presence of fluxes was demonstrated to have the same 4d manifestation (giving rise to different pairings of gauge and matter fields; we review the resulting gaugings in section \ref{em gauging}). For this reason, $e_{i A}$ and $m_i{}^A$ are sometimes referred to as geometric fluxes. \cite{KashaniPoor:2006si} emphasizes that if this procedure is to correspond to a non-linear ansatz, the expansion forms must be assumed to be moduli dependent (just as the harmonic forms on which Calabi-Yau reductions are based exhibit such dependence). \cite{KashaniPoor:2006si} then outlines which properties these forms must satisfy, given such moduli dependence, in order for the reduction of the metric sector to yield the two special geometry manifolds (the scalar manifold of the vector multiplets as well as the base of the scalar manifold of the special quaternionic manifold, the scalar manifold for the hypermultiplets) required by 4d supergravity. 

An obvious omission in this program to date is an intrinsic definition of the forms $\omega_i, \alpha_A, \beta^A$ the reduction is to be based on. For the case of nearly K\"ahler manifolds, we redress this issue by explicitly constructing a set of expansion forms in section \ref{choice of expansion forms}.

\section{Nearly K\"ahler manifolds} \label{nearlyK}
Of the many equivalent definitions of nearly K\"ahler manifolds, we choose to introduce them as $SU(3)$ structure manifolds for which the only non-vanishing torsion class is $W_1$ \cite{Alexandrov:2004cp}. The merit of this class of $SU(3)$ structure manifolds for us is that the fundamental 2- and 3-form $J$ and $\Omega$ are eigenforms, to a fixed eigenvalue determined by $W_1$, of the Laplace-Beltrami operator associated to the metric the forms determine. This hence answers the question of among which finite set of forms the expansion forms introduced in the previous section must be chosen.

\subsection{Properties and examples of nearly K\"ahler manifolds}
Nearly K\"ahler manifolds are classified by Nagy \cite{Nagy:2002}: a complete and simply connected nearly K\"ahler manifold is the Riemannian product of a K\"ahler and a strictly nearly K\"ahler (i.e. non-K\"ahler) manifold. All compact nearly K\"ahler manifolds in dimensions 2 and 4 are automatically K\"ahler. Only 4 compact strictly nearly K\"ahler manifolds are known in dimension 6, and all are homogeneous,
\ba
S^6 &\simeq& G_2/ SU(3) \,\,,\\
\IC \IP^3 &\simeq& Sp(2)/ SU(2) \times U(1) \,\,,\\
S^3 \times S^3 &\simeq& SU(2) \times SU(2) \,\,,\\
F(1,2) &\simeq& SU(3) / U(1) \times U(1) \,,
\ea
(F(1,2) is the complete flag manifold on $\IC^3$, i.e. the space of tuples $(V_1, V_2)$ of vector subspaces of dimension 1 and 2, such that $V_1 \subset V_2$). Finally, it is a theorem \cite{Butruille:2006} that this is an exhaustive list of dimension 6 homogeneous strictly nearly K\"ahler manifolds.

\subsection{Deformations of strictly nearly K\"ahler structure}
Infinitesimal deformations of nearly K\"ahler structures are studied in 
\cite{Moroianu:2006}. This analysis is somewhat orthogonal to the study in this paper, as these authors fix the normalization of $W_1$, which is the only modulus we keep in our analysis. The result of 
\cite{Moroianu:2006} is that on 6 dimensional strictly nearly K\"ahler manifolds other than $S^6$, the space of infinitesimal deformations of the nearly K\"ahler structure is isomorphic to the eigenspace of the Laplace-Beltrami operator, restricted to the space of co-closed primitive (1,1) forms, to the eigenvalue 12 (this is for $W_1 = - 2i$). These (1,1) forms parametrize the variation of $J$. It is not known whether these deformations are obstructed.

$S^6$ requires a slightly different treatment \cite{MR2322410}. Unlike all other 6d strictly nearly K\"ahler manifolds, a deformation of the nearly K\"ahler structure on $S^6$ does not necessarily involve a simultaneous deformation of the metric and the almost complex structure. In fact, for the round metric $g_0$, \cite{Moroianu:2006} determine the space of infinitesimal deformations and find that it is unobstructed, and coincides with the space of isometries of $g_0$, modded out by the isotropy group at the nearly K\"ahler structure. In other words, nearly K\"ahler deformations of $g_0$ are obtained by fixing $g_0$ and acting by its isometries on the almost complex structure.

\section{The choice of expansion forms for nearly K\"ahler manifolds} \label{choice of expansion forms}
Nearly K\"ahler manifolds have $W_1$ as their only non-vanishing torsion element,
\ba 
dJ &=& - \frac{3}{2} \im (W_1 \bar{\Omega})  \,,\\
d \Omega &=& W_1 J^2 \,.
\ea
Note that by proper choice of the phase of $\Omega$, we can choose $W_1$ to be purely imaginary. On such manifolds, both $J$ and $\Omega$ are eigenforms of the Laplacian $\triangle = d^\dagger d + d d^\dagger= -( *d*d + d*d*)$, to eigenvalue $3 |W_1|^2$.
Using $*\Omega = - i \Omega$ and $*J = \frac{1}{2} J^2$, and the fact that $W_4 =0 $ implies that $J$ is co-closed, this follows upon a straightforward calculation. Further, any manifold admitting a nearly K\"ahler structure admits a one real dimensional family of such structures, obtained by rescaling $J$, $W_1$, and $\Omega$ appropriately (note that in the mathematics literature, the normalization of $W_1$ is often fixed, thus choosing a representative of this family).

These two observations are the key ingredients in our study. 

The task is now to choose a basis of expansion eigenforms at each point in moduli space that satisfies the properties outlined in \cite{KashaniPoor:2006si}.

Within the context of Calabi-Yau like reductions reviewed in section \ref{setup}, the deformation theory of nearly K\"ahler manifolds forces us to restrict consideration to a single expansion 2-form, and 2 expansion 3-forms, yielding a 4d theory of a single vector multiplet and only the universal hypermultiplet. To see this, note that by our discussion in section \ref{nearlyK}, a deformation of nearly K\"ahler structure other than on $S^6$ is fully determined by deformations of $J$. The reduction ansatz based on the differential system (\ref{diffsys}) however necessarily yields at least twice as many  
3-forms as 2-forms. Adding degrees of freedom to $J$ hence necessarily adds degrees of freedom to $\Omega$, and leads us out of the class of nearly K\"ahler metrics.\footnote{Note that a restriction to `rigid' $\Omega$ also arose in the context of reduction of the heterotic string in \cite{Micu:2004tz}.} On $S^6$, including the isometries in our considerations would increase the number of vector fields and take us out of the context of $\cN=2$ theories in 4d. We hope to return to the question of incorporating such additional degrees of freedom in the reduction in the future.

For now, given a nearly K\"ahler structure to eigenvalue $\lambda$, we define the expansion form $\omega$ as
\ba
\omega := \frac{k}{\sqrt{\lambda}} \frac{J}{||J||} \,,
\ea
with arbitrary coefficient $k$ (below, we will see that $k = e_{10}$, with $e_{10}$ the coefficient in the expansion (\ref{diffsys}) of $d\omega$). The virtue of this definition is that it is invariant under rescaling of $J$, i.e. on the one dimensional family of nearly K\"ahler structures we are considering: on co-closed 2-forms,
\ba
\triangle_2({vJ}) = \frac{1}{v} \triangle_2(J) \,,
\ea
where the metric dependence of $\triangle_2$ is indicated in parentheses. Hence,
\ba
\triangle_2(vJ) vJ &=& \lambda_{vJ} vJ \\
&=& \frac{\lambda}{v} vJ \,,
\ea
i.e. $\lambda_{vJ} = \frac{\lambda_J}{v}$. Finally, for the norm of 2-forms,
\ba
|| \rho ||_{vJ} = \sqrt{v} || \rho||_{J} \,,
\ea
hence
\ba
||vJ||_{vJ} &=& v^{3/2} ||J||_J \,,
\ea
from which the claim follows.
By $W_4=0$, $d^\dagger J =0 $, hence $\omega$ is co-closed as well. To define a dual 4-form $\tilde{\omega}$, such that $\int \omega \wedge \tilde{\omega} = 1$, we calculate the normalization constant
\ba
g &:=& \int \omega \wedge *\omega \\
&=& \frac{k^2}{\lambda} \,,
\ea
and set 
\ba
\tilde{\omega} := \frac{1}{g} * \omega  \,.
\ea
We can define our set of expansion 3-forms via
\ba
\beta := \frac{1}{e_{10}} d\omega \,, \hspace{1cm} \alpha :=- *  \beta 
\ea 
for an arbitrary constant $e_{10}$. By
\ba
\int \alpha \wedge \beta &=& - \frac{1}{e_{10}^2} \int *d\omega \wedge  d \omega \\
&=&  \frac{k^2}{e_{10}^2} \\
&=^{\!\!\!\!{}^{!}}& 1\,,
\ea
we see that the `metric flux' $e_{10}$ merely offsets the normalization constant in the definition of $\omega$, hence has no geometric significance. In the 4d theory, shifting $k = e_{10}$ corresponds to scaling the gauge coupling constant at the expense of the normalization of the charges. There is no natural integral structure in this scheme.

The conditions on the expansion forms listed in  \cite{KashaniPoor:2006si} are easily seen to be satisfied by this set of forms: the compatibility conditions between the 2- and 3-forms reduce to $\omega \wedge d \omega = \omega \wedge * d\omega=0$ and follow from $\omega \sim J$ and the compatibility of $J$ and $\Omega$. Due to our restriction to rigid $\Omega$, the (*)ed conditions (i.e. the conditions resulting from moduli dependence of the expansion forms) on the 3-forms are trivial. The (*)ed condition on the 2-forms, $v^i \partial_j \omega_i =0$, reduces in the case of a single expansion form to constancy of the expansion form on moduli space, and this was the condition we took to motivate our definition of $\omega$ above.

The triple intersection number is obviously constant for moduli independent $\omega$. We can express this number as
\ba
\int \omega \wedge \omega \wedge \omega &=& \frac{1}{v^3} \int J \wedge J \wedge J \\
&=& 2 \frac{||J||^2}{v^3} \\
&=& \frac{2}{\lambda^{3/2} ||J||} \,.
\ea
As a consistency check, note that the last expression is indeed invariant under rescaling of $J$.

As we are parametrizing the variation of almost complex structure via deformations of $J$, the coefficients in $\Omega = Z \alpha - G \beta$ are determined in terms of $v$ (this is the analogue of having $\Omega$ fixed by its normalization in the case of Calabi-Yau manifolds with rigid complex structure). Noting that 
\ba 
d\Omega = Z\, \tilde{\omega} = W_1 J \wedge J = 2 W_1 v g \,\tilde{\omega}
\ea
and choosing a normalization of $\Omega$ such that $W_1$ is purely imaginary, $\lambda= 3 | W_1|^2 = - 3 W_1^2$, 
\ba
Z = 2 i \frac{v}{\sqrt{3 \lambda}}= 2 i \sqrt{\frac{Cv^3}{6}} \,,
\ea
and by $* \Omega = -i \Omega$,
\ba
G = -i Z = 2 \sqrt{V} \,.
\ea

We conclude this section by determining the triple intersection number for $S^6$. \cite{Alexandrov:2004cp} determines the eigenvalue $\lambda$ in terms of the Ricci scalar of the manifold,
\ba
\lambda &=& \frac{2}{5} R \,.
\ea
Note that this is the smallest eigenvalue of the Laplacian on co-closed 2-forms on $S^6$ \cite{MR0454884,VanNieuwenhuizen:1985be}.
Using $V = \frac{1}{6} C v^3$ and $\lambda = \frac{2}{Cv}$, together with $V_{S^6} = \frac{\pi^{6/2} r^6}{\Gamma(\frac{6}{2}+1)}$ and $R_{S^6} = \frac{6(6-1)}{r^2}$ yields
\ba
C_{S^6} &=& (\frac{1}{6\pi})^{\frac{3}{2}} \,.
\ea

\section{Electric-magnetic gauging and the 4d action} \label{em gauging}
Arguing from a 10d vantage point, we can put forth the following criterium for the existence of a minimum of the gauged supergravity potential: both electric and magnetic gauging must be present (we use this intuitive terminology here for convenience; this section is largely devoted to reviewing how this terminology can be made precise). The argument is simple: fluxes contribute to the energy of the field configuration via the RR kinetic terms $\sim \int F_n \wedge *F_n$. A simple counting of powers of the metric establishes that a rescaling of the metric $g_{mn} \mapsto \lambda^2 g_{mn}$ results in a rescaling of this contribution by $\lambda^{2(3-n)}$. The contributions of both $F_0$ and $F_2$ hence scale inversely, as compared to the contributions of $F_4$ and $F_6$, under rescaling of the size of the compactification manifold. Therefore, in order for the manifold to be stabilized at finite radius, fluxes for $n$ both larger and smaller than $3$ must be present. Stabilization at finite radius translates into a minimum of the 4d potential at finite K\"ahler moduli. Now, $n=3$ is also the bound that determines whether fluxes result in electric or magnetic gauging in the 4d effective theory
\cite{Louis:2002ny}, thus completing the argument.\footnote{Note that there is an interesting parallel here with the no-go theorem \cite{Cecotti:1984wn, Mayr:2000hh,Louis:2002vy} regarding partial breaking of supersymmetry in $\cN=2$ supergravity for Minkowski solutions. The observation there is that partial supersymmetry breaking in the conventional framework of $\cN=2$ supergravity is not possible unless a degenerate choice of the symplectic vector $(X^I, F_I)$ is made. Under a symplectic rotation, the degeneracy of this choice can be undone, but only at the expense of generating magnetic gaugings.}
While simultaneous electric and magnetic gauging is possible at the level of the equations of motion, it cannot naively be implemented in a local action.
The most familiar formulation of gauged $\cN=2$ supergravity \cite{Andrianopoli:1996cm} in terms of vector and hypermultiplets is hence not sufficient for our purposes. Luckily, starting with \cite{Louis:2002ny}, we have learned how to implement these equations of motion by including tensor multiplets in the $\cN=2$ action \cite{Theis:2003jj,DallAgata:2003yr,DAuria:2004yi,DAuria:2007ay} . In this section, we wish to review this development and its relation to the very intuitive `symplectic completion' \cite{Michelson:1996pn} of the standard formalism \cite{Andrianopoli:1996cm}, in particular the elegant packaging of compactification data in terms of symplectically completed killing prepotentials as worked out in \cite{Grana:2005ny,Grana:2006hr}.

\subsection{Quaternionic geometry of the hypermultiplet sector}
A quaternionic K\"ahler manifold of dimension $4n$ is by definition an oriented Riemannian manifold with holonomy group contained in $Sp(1) \otimes Sp(n)$. The quaternionic metrics that arise at tree level in Calabi-Yau compactifications were worked out in \cite{Ferrara:1989ik}. These are coordinatized by the dilaton $\phi$, axion $a$ (dual to $B$), the complex structure moduli  $z_i$ (in the case of IIA) of the Calabi-Yau, and the axions $\xi^A, \txi_A$ stemming from RR fields. They are termed special quaternionic metrics, as the RR axions are fibered over the special geometry directions coordinatized by the complex structure moduli. The metric takes the explicit form\footnote{The normalization here differs slightly from the one in \cite{Louis:2002ny}, which took an unconventional normalization of the RR field strengths as a starting point of the reduction, see also \cite{Bohm:1999uk}. This choice only becomes relevant when comparing 4d and 10d solutions.}
\ban
h_{uv} dq^u \otimes dq^v &=& g_{i \bar{\jmath}} dz^i \otimes d\bar{z}^{\bar{\jmath}} + d \phi \otimes d \phi \nonumber \\ & & + \frac{e^{4\phi}}{4}\left[da + \frac{1}{2}(\txi_A d\xi^A - \xi^A d\txi_A)\right]\otimes \left[da + \frac{1}{2}( \txi_A d\xi^A - \xi^A d\txi_A)\right] \nonumber \\
& &-\frac{e^{2\phi}}{4} (\im \cM^{-1})^{AB}\left[d\txi_A  +\cM_{AC}d\xi^C \right] \otimes   \left[d\txi_B  +\overline{\cM}_{BD}d\xi^D \right] \,, \label{qmetric}
\ean
where $g_{i \bar{\jmath}}$ and $\cM$ are determined by special geometry data ($\cM$ is the mirror of the gauge coupling matrix $\cN$, see appendix \ref{specialK}). With regard to the metric $\frac{1}{2} \epsilon_{ab} \epsilon_{AB}$, \cite{Ferrara:1989ik} introduces the vielbein
\ban
\cU =  \left(  \begin{matrix} 
      u & e & -\bar{v} & -\bar{E} \\
      v & E & \bar{u} & \bar{e} \\
   \end{matrix} \right)  \label{vielbein}
\ean
on the complexified tangent space of the manifold, on which the two factors of the holonomy act on the left, right respectively. Of the entries in this matrix, only $u$, $v$ will be relevant for us in the following,
\ba
u&=& -\frac{i}{\sqrt{2}} e^{\frac{K}{2}+ \phi} Z^A (d\txi_A + \cM_{AB} d\xi^B) \,, \\
v&=& d\phi - i\frac{e^{2 \phi}}{2} \left(da + \frac{1}{2}(\txi_A d\xi^A - \xi_A d \txi^A)\right)  \,.
\ea 
The connection of the metric decomposes according to the $Sp(1) \otimes Sp(n)$ factorization of the holonomy,
\ban
d\cU= \omega \wedge \cU - \cU \wedge \Delta \,. \label{connection}
\ean
The relevant quantity for us is the $Sp(1)$ connection $\omega = \frac{i}{2}\omega^x (\epsilon \sigma_x \epsilon^{-1})$, with $\sigma_x$ the Pauli matrix basis of $\mathfrak{su}(2)$, given by
\ba
\omega^1 = i (\bar{u} - u)\,\,,\,\,\, \omega^2 = -(u + \bar{u}) \,,\\
\omega^3 = \frac{i}{2} (v - \bar{v}) + \ldots \,,
\ea
the $\ldots$ subsuming directions in the special geometry base.

Quaternionic K\"ahler manifolds $M$ are a local version of hyperk\"ahler manifolds, in that they locally exhibit a triplet of almost complex structures $J^x$ satisfying the quaternionic algebra. For our purposes, it is convenient to phrase this structure in terms of an $SU(2)$ principal bundle ${\cal V}$ on $M$, with connection the $\omega$ introduced in (\ref{connection}). Locally, the bundle ${\cal V}\otimes \Lambda^2 T^*M$ is trivialized by a triplet of flat sections $K^x$, $x=1,2,3$,
\ba
\nabla K^x = dK^x + \epsilon^{xyz} \omega^y \wedge K^z = 0 \,,
\ea
related to the almost complex structure via $K^x(\cdot,\cdot) = h(J^x \cdot, \cdot)$. The $K^x$ are called hyperk\"ahler forms, though unlike the hyperk\"ahler case, they are not global objects. 

The quaternionic metric (\ref{qmetric}) has a set of isometries given by
\ban
{\bf k_c}= {\bf \partial_a} \,\,\,,\,\,\,\,\, {\bf k^A} = - \frac{1}{2} \xi^A {\bf \partial_a} + {\bf \partial_{\txi_A}} \,\,\,,\,\,\,\,\, {\bf k_A} = \frac{1}{2} \txi_A {\bf \partial_a} + {\bf \partial_{\xi^A}} \,.  \label{isometries}
\ean
These span a Heisenberg algebra,
\ba
[{\bf k^A}, {\bf k_B}] &=&  \delta^A_B\, {\bf k_c}  \,,
\ea
with ${\bf k_c}$ as central element. Quaternionic K\"ahler manifolds permit a generalization of the moment map construction \cite{Galicki:1986ja}: despite not being globally defined, the hyperk\"ahler forms $K^x$ can be used to introduce moment maps for isometries ${\bf k}$ via
\ba
\nabla \cP^x_{\bf k} &=& - \iota_{\bf k} K^x \,,
\ea
where $\iota$ signifies contraction. The moment maps $\cP^x$ are called killing prepotentials. Note that due to the local nature of $K^x$, we are forced to introduce a triplet of moment maps, which are local sections of ${\cal V}$, and that the covariant derivative appears on the LHS of the moment map equation, rather than the more familiar straight differential. Due to this, the definition of the moment maps are possible for isometries which preserve the Hyperk\"ahler forms only up to a so-called $SU(2)$ compensator $W^z_k$,
\ba
{\cal L}_{\bf k} K^x &=& \epsilon^{xyz} K^y W^z_{\bf k} \,.
\ea
It is a pleasant surprise that the seeming complication of having a non-trivial $SU(2)$ bundle allows for an algebraic, rather than a differential, relation between the killing vectors, the $SU(2)$ compensator, and the killing prepotentials,
\ba
\epsilon^{xyz} K^y W^z_{\bf k} &=& - \epsilon^{xyz} ( \iota_{\bf k} \omega^y - \cP^y_{\bf k}) K^z  \,.
\ea
As the isometries (\ref{isometries}) of the metric we consider in fact preserve the quaternionic structure without the need for a compensator \cite{Michelson:1996pn}, this relation becomes
\ban
\cP_{\bf k}^x &=& k^u \omega_u^x \,.   \label{simple prepotentials}
\ean

In the context of flux compactifications on SU(3) structure manifolds, which of the isometries (\ref{isometries}) is gauged, and by which vector, is encoded in the RR and NS background field strengths,
\ba
F_0 = m \,\,,\,\,\, F_2 = m^i \omega_i &,& F_4 = e_i \tomega^i \,\,,\,\,\, F_6 = e\, \frac{vol}{V} \,, \\
H &=& p^A \alpha_A - q_A \beta^A \,,
\ea
as well as the integers appearing in the differential system (\ref{diffsys}) specified by the expansion forms. Which integers correspond to which gauging \cite{Louis:2002ny,Gurrieri:2002wz,DAuria:2004tr} is easy to remember based on the index structure: if we denote the isometry gauged by the $i^{th}$ vector multiplet as ${\bf k_i}$, and by the graviphoton as ${\bf k_0}$, then
\ban
{\bf k_0} &=& p^A {\bf k_A} + q_A {\bf k^A} -  e {\bf k_c} \,, \nonumber\\
{\bf k_i} &=& e_{iA} {\bf k^A} + m_i^A {\bf k_A} - e_i {\bf k_c} \,, \nonumber\\
{\bf \tilde{k}^0} &=& m {\bf k_c} \,, \nonumber \\
{\bf \tilde{k}^i} &=& m^i {\bf k_c} \,, \nonumber \\ \label{gauged isometries}
\ean
(note the necessity to distinguish between ${\bf k_0}$, the killing vector gauged by the graviphoton, and ${\bf k_{A=0}}$). The relevance of the tilded killing vectors is that these are gauged magnetically. Hence, whenever we consider a reduction in the presence of fluxes $F,G$ such that $\int_{X_6} F \wedge G \neq 0$, the non-gravitational sector of the $\cN=2$ 4d action cannot be described, as in \cite{Andrianopoli:1996cm}, purely in terms of vector and hypermultiplets \cite{Louis:2002ny}: at the level of the equations of motion, such 10d backgrounds give rise to 4d hyperscalars that are charged both electrically and magnetically under the same gauge field.

\subsection{Dualizing scalars to tensors to accommodate magnetic charges} \label{tensor multiplets}
The observation that considering compactifications in the presence of $F_0, F_2, F_4, F_6$ flux yields scalar fields charged electrically and magnetically under the same gauge field is first made in \cite{Louis:2002ny}. The resolution to the problem of capturing this setup in a local action is also presented in \cite{Louis:2002ny}: the 4d action can be formulated by dualizing the culprit doubly charged scalars to  tensor fields.\footnote{The authors of \cite{deWit:2005ub} take a different approach to this problem: they introduce both electric and magnetic gauge potentials in the action, together with gauge symmetries tied to tensor fields to compensate for the surplus in degrees of freedom. It is an intriguing question whether such an action can be obtained upon reduction, taking the formulation of the 10d supergravity action developed in \cite{Belov:2006xj} as a starting point.} More specifically, one can gauge the isometry ${\bf \partial_a}$ of the conventional $\cN=2$ action electrically. As this does not break the shift symmetry of the action in $a$, one can next dualize $a$ to a tensor $B$. Finally, the obtained action can be deformed by adding couplings between the gauge fields and the tensor $B$ parametrized by the erstwhile magnetic charges of $a$,
\ba
F^I &=& d A^I + m^I B \,,
\ea
with $I=(0,i)$. \cite{Louis:2002ny} demonstrates that precisely this action is obtained upon reduction, by refraining from the conventional dualization to a scalar of the spacetime components of the NSNS $B$-field. Finally, the authors of that paper demonstrate that the potential they obtain from the reduction is precisely the one that was originally suggested in \cite{Michelson:1996pn}, the naive symplectic completion of the potential presented in \cite{Andrianopoli:1996cm},
\ba
V &=& 4 e^K h_{uv} ( X^I k_I^u - \tilde{k}^{uI} F_I) ( \bar{X}^I k_I^u - \tilde{k}^{uI} \bar{F_I}) \\
& & -\left[ \frac{1}{2} (\im \cN)^{-1 \,IJ} + 4 e^K X^I \bar{X}^J \right] ( \cP_I^x - \tP^{Kx} \cN_{KI}) ( \cP_J^x - \tP^{Lx} \bar{\cN}_{LJ}) \,.
\ea
Note that this potential does not depend on the scalar fields being dualized.

The results of \cite{Louis:2002ny} are not quite sufficient for our purposes, as the deformation (\ref{diffsys}) we are considering is to  gauge the isometry ${\bf k^A}$ in addition to ${\bf k_c}$, but this isometry is broken upon dualizing $a$ to a tensor. The resolution to this problem is a simple coordinate redefinition \cite{DAuria:2007ay}. By replacing $a$ by $\hat{a}$, 
\ba
\hat{a} = a - \frac{1}{2} \xi^A \txi_A  \,,
\ea
the isometries ${\bf k^A}$ become simple shift symmetries of $\txi_A$ (more generally, we can of course also choose the definiton of $\hat{a}$ to allow the gauging of ${\bf k_A}$ for some $A$). After gauging both ${\bf \partial_{\tilde{a}}}$ and ${\bf \partial_{\txi_A}}$ electrically, they can hence be dualized to tensors, and the `magnetic' deformations introduced as before.

Since \cite{Louis:2002ny}, the modifications to the conventional $\cN=2$ gauged supergravity action 
\cite{Andrianopoli:1996cm} in the presence of tensor multiplets have been extensively studied \cite{Theis:2003jj,DallAgata:2003yr, DAuria:2004yi,DAuria:2007ay}. In particular, \cite{DAuria:2004yi} derives the full $\cN=2$ action together with its supersymmetry variations in the generality we require.\footnote{\cite{DAuria:2004yi} first dualizes a set of scalar fields to tensors and then deform the action electrically and magnetically, with deformation parameters $e^I_\Lambda$, $m^{I\Lambda}$. For the situation we are considering, this is equivalent to first gauging isometries, as parametrized by charges $e^I_\Lambda$, dualizing, and then deforming magnetically \cite{DAuria:2007ay}.} That the resulting action is the one one obtains upon reduction from IIA has not been demonstrated completely yet, though many components of a general proof are in place \cite{Louis:2002ny,Gurrieri:2002wz,Gurrieri:2002iw,DAuria:2004tr,Kachru:2004jr,Grana:2005ny,Louis:2006kb,KashaniPoor:2006si,Grana:2006hr}. Rather than compactifying the bosonic action, the authors of \cite{Grana:2005ny,Grana:2006hr} take the gravitino transformation properties as a starting point. They derive the 4d gravitino mass matrix $S_{ab}$, and utilize the relation \cite{Andrianopoli:1996cm}
\ban
S_{ab} = \frac{i}{2} e^{\frac{1}{2} K} \sigma^x_{ab} \cP^x  \label{gravitino mass in terms of prepotentials}
\ean
to obtain expressions for the symplectically completed quaternionic killing prepotentials $\cP^x$,
\ban
\cP^x &=& \cP^x_I X^I - \tilde{\cP}^{x I} F_I \,. \label{gkp}
\ean
In the remaining part of this section, we will verify in a straightforward calculation that the potential and the supersymmetry transformations of the fermions as worked out in \cite{DAuria:2004yi} can indeed be expressed in terms of the generalized killing prepotentials (\ref{gkp}).

In \cite{DAuria:2004yi}, the fermion independent terms that appear in the supersymmetry transformations of the fermions 
\ba
\delta \psi_{a \mu} &=& \nabla_\mu \epsilon_a - S_{ab} e^{-\phi} \gamma_\mu \epsilon^b \,,\\
\delta \zeta_\alpha &=& N^a_\alpha \epsilon_a \,, \\
\delta \lambda^{ia} &=& W^{iab}\epsilon_b \,
\ea
are 
\ba
S_{ab} &=& \frac{i}{2} e^{\frac{K}{2}} \sigma^x_{ab} \omega_\Lambda^x ( e_I^\Lambda X^I - m^{\Lambda I} F_I ) \,, \\
W^{iab} &=& i g^{i \bar{\jmath}} \sigma^{x\,ab} \omega_\Lambda^x (e_I^\Lambda \bar{f}_{\bar{\jmath}}^I - m^{\Lambda I} \bar{h}_{\bar{\jmath} I}) \,,\\
N_\alpha^a &=& 2\,e^{\frac{K}{2}} {\cal U}_\Lambda{}^a{}_\alpha (e_I^\Lambda X^I - m^{\Lambda I} F_I) \,.
\ea
$\Lambda$ here is a tensor multiplet index. It takes on the values corresponding to the two dualized quaternionic directions $\hat{a}$, $\txi$.\footnote{As we will not be considering magnetic charges for $\txi$, we could equally well keep the scalar variable and gauge its shift symmetry, see  previous footnote.} $\epsilon_a$ is a section of the bundle ${\cal A} \otimes {\cal S_+}X$, where ${\cal A}$ is the associated bundle (for the fundamental representation) to the $SU(2)$ principal bundle ${\cal V}$ introduced in the previous subsection, and ${\cal S_+}X$ denotes the positive chirality spin bundle to the 4d spacetime manifold (i.e. it is a spacetime spinor with $SU(2)$ R-symmetry index). $\epsilon^a$ has opposite chirality. The special K\"ahler ingredients in the above equations are explained in appendix \ref{specialK}.

By considering cases which do not require dualization, we arrive, comparing to (\ref{gauged isometries}), at the following identification of parameters, $e^{\hat{a}}_0 = - e_0$, $e^{\hat{a}}_i = - e_i$, $e^{\txi_A}_0 = q_A$, $e^{\txi_A}_i = e_{iA}$, $m_0^{\hat{a}} =  m_0$, $m^{\hat{a}}_i =  m_i$.

Since we are not rotating the vielbein $\cU$ by passing from $a$ to $\hat{a}$, the connection transforms in a simple fashion. Explicitly, $\omega^1$ and $\omega^2$ remain unchanged, while
\ba
\omega^3 &=& \frac{e^{2\phi}}{2} (d \hat{a} -  \xi^A d \txi_A)  + \ldots \,,
\ea
and compared to (\ref{gravitino mass in terms of prepotentials}), using (\ref{simple prepotentials}), we obtain the identification $\cP^x_I = \omega_\Lambda^x e_I^\Lambda$, $\tilde{\cP}^{x\,I} = \omega_\Lambda^x m^{\Lambda I}$.

\subsection{Our 4d theory}
Given our choice of expansion ansatz as described in section \ref{choice of expansion forms}, the internal components of $H$, $G_2$ and $G_4$ are necessarily cohomologically trivial,
\ba
H^{int} = b \,d \omega = b \beta \,\,\,,\,\,\,\,\, G_2^{int} =0 \,\,\,,\,\,\,\,\,G_4^{int} = \xi \,d \alpha + \tilde{\xi}\, d \beta = \xi\, \tilde{\omega} \,.
\ea
The only honest fluxes we have access to are 
\ba
G_0^{int} = m \,\,\,\,\,\,, \hspace{1cm} G_6^{int} = e \,\frac{vol}{V} \,.
\ea
We can read off the isometries being gauged (in the sense explained in the previous subsection) from 
(\ref{gauged isometries}). The generalized killing prepotentials, which will feature prominently in the next section, are
\ba
\cP^1 &=& 0\\
\cP^2 &=&   -  e^\phi t \,,\\
 \cP^3  & =& -   \frac{e^{2\phi}}{2} \big[ X^1  e_{1\o}\xi  +
  X^0 e_0 + F_0 m^0 )\big] \\
&=& -   \frac{e^{2\phi}}{2}
\big[ t e_{1\o}\xi  +
   e + \frac{1}{6}C t^3 m )\big] \,.
\ea

The RR field strengths discussed so far satisfy the Bianchi identities $(d - H_{flux}) G =0$. As we are considering the case without $H$-flux, all $G$ must be closed, as realized by our ansatz. It is often convenient to also work with an alternative basis of RR fields, defined via
\ba
F &=& e^B G \,.
\ea
These satisfy the Bianchi identities $(d-H)F=0$. The constraints on the RR fields coming from SUSY variations are more succinctly formulated in terms of the $F$ basis, while the relations between charges and fluxes is more direct in the $G$ basis.

In components, the two bases are related by 
\ba
F_0 &=& G_0 = m =f_0\,, \\
F_2 &=& G_2 + B \wedge G_0 = b m \, \omega =f_2 \,\omega\,,\\
F_4 &=& G_4 + B \wedge G_2 + \frac{1}{2} B \wedge B \wedge G_0 \\
&=& (\xi + \frac{1}{2} C b^2 m) \tilde{\omega} = f_4 \,\tilde{\omega}\,, \\
F_6 &=& G_6 + B \wedge G_4 + \frac{1}{3!} B^3 \wedge G_0 \\
&=& (e + b \xi + \frac{Cb^3 m}{6}) \frac{1}{C} \,\omega \wedge \omega \wedge \omega = f_6\, \frac{vol}{V} \,.
\ea

\section{Lifting supersymmetric 4d solutions} \label{lifting}
In this section, we demonstrate that the supersymmetric solutions of our 4 dimensional effective action lift to the 10d nearly K\"ahler solutions which have been derived from a 10d point of view in \cite{Behrndt:2004km, Behrndt:2004mj, Lust:2004ig, Grana:2006kf}. Note that a similar goal is pursued in \cite{House:2005yc}, but with a focus on an $\cN=1$ formulation in 4d.

In this section, we first apply the general analysis of \cite{Cassani:2007pq} to our setup. To compare to the 10d analysis of \cite{Grana:2006kf}, we solve the $\cN=1$ equations arising from setting the 4d fermion variations to 0 to express the fluxes in terms of essentially the 4d cosmological constant. As expected from the analysis of \cite{Cassani:2007pq}, we find agreement with the 10d analysis of \cite{Grana:2006kf}. We then re-express our results in a more natural way with regard to the 4d theory, by expressing all moduli fields in terms of the $G_0$ and $G_6$ flux parameter. Finally, we verify explicitly that the solutions to the $\cN=1$ constraints indeed minimize the 4d potential.

\subsection{Solving the 4d $\cN=1$ constraints}
The calculations in this subsection are a specialization of the analysis that appears in section 4 of \cite{Cassani:2007pq}. The starting point is requiring the vanishing of the supersymmetry transformations of the gravitino $\psi_{A \mu}$, hyperinos $\zeta_\alpha$, and gauginos $\lambda^{iA}$, 
\ban
\delta_\epsilon \psi_{a \mu} &=& 0 \,, \nonumber\\
\delta_\epsilon \zeta_\alpha &=& 0 \,, \nonumber\\
\delta_\epsilon \lambda^{ia} &=& 0 \,.  \label{fermion variations}
\ean
As noted in subsection \ref{tensor multiplets}, $\epsilon_a$ is a section of ${\cal A} \otimes {\cal S_+}X$. Choosing a local trivialization of ${\cal A}$ and a section $\epsilon$ of ${\cal S_+}X$ satisfying the killing spinor equation $\nabla_\mu \epsilon = \frac{1}{2} \mu \gamma_\mu \epsilon^*$, we can locally set
\ba
\left(\begin{array}{c} \epsilon_1 \\ \epsilon_2 \\ \end{array}\right) &=& \left(\begin{array}{c} a \\ b \\ \end{array}\right) \epsilon \,,
\ea
with the normalization $|a|^2 + |b|^2 = 1$.

Let's first deal with the factors $a$ and $b$. It is straightforward to check \cite{Cassani:2007pq} that the hyperino equations yield
\ba
\bar{a} (\cP^1 - i \cP^2) - 2 \bar{b} \cP^3 &=& 0  \,,\\
\bar{b} (\cP^1 - i \cP^2) + 2 \bar{a} \cP^3 &=& 0 \,,
\ea
while the gravitino equations are equivalent to
\ba
\bar{a} (\cP^1 - i \cP^2) -  \bar{b} \cP^3 &=& -i e^{-\frac{K}{2}+\phi} a \mu \,, \\
\bar{b} (\cP^1 - i \cP^2) + a \cP^3 &=&  i e^{-\frac{K}{2}+\phi} b \mu \,.
\ea
Together, these equations imply \cite{Cassani:2007pq} $(|a|^2 - |b|^2) \mu = 0$. Since $\mu =0$, i.e. a Minkowski vacuum, is not compatible with $e \neq 0, m \neq 0$, we can conclude
\ba
|a|^2 - |b|^2 &=& 0 \,.
\ea
Next, imposing this condition on the phases, the gaugino variation yields \cite{Cassani:2007pq}
\ba
\re \left( \bar{a}b(\cP^1_I - i \cP^2_I) \right) &=& 0 \,.
\ea
Since we have $\cP^1=0$ and $\cP^2_I \in \IR$, this forces $\bar{a}b \in \IR$, hence
\ba
a = b \,,
\ea
and the gravitino equations take the simple form
\ban
\frac{1}{2} \cP^2 = i\cP^3 = 2 a^2 \mu e^{\phi - \frac{K}{2}} \,. \label{grequation}
\ean

With these conditions in place, let us now turn to solving the equations (\ref{fermion variations}). It proves computationally convenient \cite{Cassani:2007pq} and facilitates comparison to the 10d literature \cite{Behrndt:2004km,Behrndt:2004mj,Lust:2004ig,Grana:2006kf}, to first determine the solution to the $\cN=1$ constraints in terms of the parameter $\mu$. The gaugino equation yields
\ba
\sigma_x^{AB} n_B \left( (\im \cN)^{-1 IJ} (\cP^x_J - \cN_{JK} \tP^{xK}) + 2 e^K \bar{X}^I \cP^x \right) = 0  \,.
\ea
Upon utilizing the gravitino equations (\ref{grequation}), this is \cite{Cassani:2007pq}
\ba
- (\im \cN)^{-1\,IJ} \cP_J^2 - i (\im \cN)^{-1\, IJ} (\cP_J^3 - \cN_{JK} \tilde{\cP}^{3\,K}) &=& 12 e^{\frac{K}{2}+ \phi} a^2 \mu \bar{X}^I \,.
\ea
This equation evaluates to
\ba -\frac{1}{2V}
 \left(\begin{array}{c} 
  i e^{2\phi} f_6 +  \frac{1}{6} e^{2 \phi} Cmv^3 + 2 b e^\phi \\
  \frac{i}{3} e^{2\phi} v^2 f_4 + i e^{2\phi} b  f_6 +
\frac{2}{3} e^\phi (v^2+3 b^2) \end{array} \right) &=& \frac{3 \sqrt{2}}{\sqrt{V}} e^\phi \tilde{\mu} \left(\begin{array}{c} 1 \\ b-iv \end{array} \right) \,,
\ea
where for convenience, we have absorbed the factor $a$ by defining $\tilde{\mu} = a^2 \mu$. Separating into real and imaginary parts, we obtain
\ba
f_6 &=& 6 - \sqrt{2V} e^{-\phi} \tilde{\mu}_I \,, \\
v f_4 &=& 18 \sqrt{2V} e^{-\phi} \tilde{\mu}_R \,,
\ea
and
\ban
f_0 &=& -(\frac{2b}{3 \sqrt{V}} + 6\sqrt{2} \tmu_R) \frac{e^{-\phi}}{\sqrt{V}} \,, \\
v^2 + 3 b^2 &=& 9 \sqrt{2V} (\tmu_R b + \tmu_I v) \,.  \label{fixes b}
\ean
The equation
\ba
\cP^2 =4 a^2 \mu e^{\phi - \frac{K}{2}} 
\ea
yields
\ban
\tmu_R = - \frac{b}{8 \sqrt{2V}} \,\,,\,\,\,\, \tmu_I = -\frac{v}{8\sqrt{2V}} \,, \label{mu}
\ean
and with the above, $\cP^2 = 2i \cP^3$ is then identically satisfied.
Plugging into (\ref{fixes b}), we finally obtain
\ba
b^2 = \frac{1}{15}v^2 \,.
\ea

Introducing the 10 dimensional dilaton via $e^{-\phi_{10}} = \frac{e^{-\phi}}{\sqrt{V}} $, we can now summarize the above findings,
\ba
W_1 &=& i \sqrt{\frac{\lambda}{3}} = -\frac{8i}{3} \sqrt{2} \tmu_I \,, \\
H&=& b\, d\omega = 4 \sqrt{2} \tmu_R \re \Omega \,, \\
F_0 &=& 10 \sqrt{2} \tmu_R e^{-\phi_{10}}\,,\\
F_2 &=& f_0 b\, \omega = \frac{2\sqrt{2}}{3} e^{-\phi_{10}} \tmu_I J \,,\\
F_4 &=&   3 \sqrt{2} e^{-\phi_{10}} \tmu_R J \wedge J \,, \\
F_6 &=&  - \sqrt{2} e^{-\phi_{10}} \tmu_I J \wedge J \wedge J \,,
\ea
where we have used $\tomega = \frac{1}{C} \,\omega \wedge \omega$, and $\frac{vol}{V} = \frac{1}{C}\, \omega \wedge \omega \wedge \omega$.
As expected from the general analysis of \cite{Cassani:2007pq}, we have been able to reproduce the results in particular of \cite{Grana:2006kf} from a 4d calculation. A comment is in order regarding the warp factor. Our reduction ansatz assumes a constant warp factor. To be precise, the expression `constant warp factor' has no invariant meeting, as such a factor can always be absorbed in the metric. As such, the factor $A$ which appears in \cite{Grana:2006kf} is naturally incorporated in $\mu$; indeed $ e^{2A} ds^2_{AdS}(\Lambda) = ds^2_{AdS}(e^{-2A} \Lambda)$, where $\Lambda \sim |\mu|^2$. 

From the point of view of the 4d theory, it is more natural to express the 4 dimensional fields (traditionally called moduli, though of course, they are not) $v, b, \phi, \xi$ in terms of the flux parameters $m$ and $e$. Reorganizing the above equations, we obtain
\ban
v_s^3 &=& \frac{9}{16} \sqrt{15} \frac{e}{Cm} \,, \nn \\
b_s &=& -\frac{1}{\sqrt{15}}v_s \,, \nn \\
\xi_s &=& \frac{4}{15} C m v_s^2 \,, \nn \\
e^{ \phi_s} &=& \frac{\sqrt{15}}{ 2 C m v_s^2} \,, \nn\\
\label{solutions}
\ean
with $\tilde{\mu}_s$ given by evaluating (\ref{mu}) on this supersymmetric field configuration. 

\subsection{Minimizing the 4d potential}
We would now like to demonstrate that the solutions (\ref{solutions}) to the $\cN=1$ constraints satisfy the 4d equations of motion, i.e. minimize the 4d potential. We first pursue a general approach as outlined in \cite{Cecotti:1984wn} and particularly \cite{Elvang:2007ba}. The starting point is the supersymmetry variation of the action. This vanishes order by order in the fermion fields. Consider a bosonic supersymmetric field configuration $(\Phi_s, \Psi_s=0)$, with $\Phi=(\phi_i)$ collectively denoting all bosonic fields, and analogously $\Psi=(\psi_i)$ the fermions. Now focus on the supersymmetric variation of the action to first order in the fermions and evaluate this on $\Phi_s$, keeping the fermionic fields general. By definition of $\Phi_s$, we obtain
\ba
(\delta_{SUSY} I[\Phi_s])_{\rm 1st\, order} = \int \sum_i \frac{\delta L}{\delta \phi_i}[\Phi_s,\Psi=0] \,\delta \phi_i[\Phi_s,\Psi]= 0 \,.
\ea 
Two obstructions separate us from concluding that the field configuration $\Phi_s$ satisfies the equations of motion: to conclude that the integrand vanishes, we must rule out total derivative terms, and to then conclude that each summand vanishes separately, we must ensure that the variations $\delta \phi_i[\Phi_s, \Psi]$ are linearly independent in the vector space spanned by the fermions. 

Let us now apply these arguments to our setup. Since we are only considering constant field configurations, the first condition is satisfied. The second must be checked explicitly. Since the potential does not depend on the dualized scalars, we only need to consider the variations of the scalars in the vector multiplet and the two remaining scalars (after dualization) in the hypermultiplet. These variations are
\cite{Andrianopoli:1996cm,DAuria:2004yi}
\ba
\delta z^i &=& \bar{\lambda}^{ia} \epsilon_a \,, \\
\delta q^u &=& {\cal U}^u_{a \alpha} (\bar{\zeta}^\alpha \epsilon^a + \epsilon^{\alpha \beta} \epsilon^{ab} \bar{\zeta}_\beta \epsilon_b) \,,
\ea
with ${\cal U}^u_{a \alpha}$ the inverse of the vielbein introduced in (\ref{vielbein}). Note that ${\cal U}$ generally satisfies the reality constraint
\ba
({\cal U}^u_{a \alpha})^* &=& \epsilon_{ab} \epsilon_{\alpha \beta} {\cal U}^u_{b \beta}  \,,
\ea
as can be explicitly verified for (\ref{vielbein}). Together with
\ba
(\bar{\zeta}^\alpha \epsilon^a)^\dagger &=& \bar{\zeta}_\alpha \epsilon_a \,,
\ea
this guarantees the reality of $\delta q^u$. 

The variations $\delta z^i$ are clearly independent. With $\epsilon^1 = \epsilon^2$, the variations of $\delta q^u$ for $u=\phi,\xi$ are 
\ba
\delta \phi &=& \re( \bzeta^1 \epsilon^1) - \re (\bzeta^2 \epsilon^1) \,,\\
\delta \xi &=& - 2 e^{-\phi_s} \left[ (\im (\bzeta^1 \epsilon^1)- \im (\bzeta^2 \epsilon^1) \right] \,.
\ea
These are likewise independent. Hence, our supersymmetric field configuration is guaranteed to be a solution of the equations of motion.

As a check on this reasoning, we now proceed to evaluate the potential explicitly and check that it is minimized by our solution.

The potential determined in \cite{DAuria:2004yi} can be expressed in the form
\ba
V &=& 4 e^K h_{uv} ( X^I k_I^u - \tilde{k}^{uI} F_I) ( \bar{X}^I k_I^u - \tilde{k}^{uI} \bar{F_I}) \\
& & -\left[ \frac{1}{2} (\im \cN)^{-1 \,IJ} + 4 e^K X^I \bar{X}^J \right] ( \cP_I^x - \tP^{Kx} \cN_{KI}) ( \cP_J^x - \tP^{Lx} \bar{\cN}_{LJ}) \,,
\ea
which is the naive symplectic completion of the potential presented in 
\cite{Andrianopoli:1996cm}, and was first proposed in \cite{Michelson:1996pn}.

The explicit form of the potential in our setup is
\ban
V &=& \frac{1}{4 C v^3} \big[ e^{4 \phi} (3 e^2 + 6 e b \xi + C e m b^3 + \frac{1}{12} C^2 m^2 (v^2 + b^2)^3 + \nonumber \\
& &+  Cm b^2 ( v^2 + b^2) \xi + (v^2 +3 b^2)\xi^2 ) +  e^{2 \phi} (-5 v^2 + 3 b^2) \big] \,.  \nonumber \\ \label{potential}
\ean
Note that setting $m=0$ removes all terms that increase with increasing $v$, as predicted by the scaling argument presented in section \ref{em gauging}.

Plugging the solution (\ref{solutions}) of the $\cN=1$ constraints into (\ref{potential}), we obtain
\ba
V(v_s,b_s,\xi_s,\phi_s) &=& -3 e^{2 \phi_s} |\mu_s|^2 \,,
\ea
as required by the Ward identities relating the $\cN=2$ scalar potential to the squares of the fermion variations \cite{Cecotti:1984wn,DAuria:1990fj,DallAgata:2003yr},
\ba
\delta^a_b V &=& -12 \bar{S}^{ca} S_{cb} + g_{i \bar{\jmath}} W^{i\,ca}W^{\bar{\jmath}}_{cb} + 2 N^A_\alpha N^\alpha_B  \,.
\ea

By our reasoning above, the solution $\{v_s, b_s, \xi_s, \phi_s \}$ to the $\cN=1$ constraints should also extremize the potential. Given the explicit form of the potential (\ref{potential}), it is easy to check that this is indeed the case, and that the extremum is a minimum.\footnote{To check that the extremum is a minimum, we ascertain that the determinant of the hessian is non-zero for any value of $C,e,m$, then verify numerically that all eigenvalues are positive for a fixed (arbitrary) choice of these constants.}

\section*{Acknowledgements}
We would like to thank Ruben Minasian for many insightful comments and suggestions. We also gratefully acknowledge useful conversations with Jan de Boer, Sheer El-Showk, Thomas Grimm, Andrei Micu, Kostas Skenderis, and Oscar Varela, and thank Uwe Semmelmann for helpful explanations of his work.

We would like to thank the Aspen Institute for Physics and the 2007 Simons Workshop, where part of this work was performed. 

This work was supported by Stichting FOM. 

\appendix

\section{Special geometry} \label{specialK}
We here collect some formulae for the special geometry sector for convenience.

In terms of a holomorphic prepotential $F$, the K\"ahler potential is given by
\ba
e^{-K}  &=& i (\bar{X}^I F_I - X^I \bar{F}_I) \,,
\ea
where $F_I = \partial_I F$. In the vector multiplet sector,
\ba
e^{-K} = 8 V = \frac{1}{6} \int J \wedge J \wedge J \,.
\ea
The metric here evaluates to 
\ba
G_{ij} &=& \frac{1}{4V} \int \omega_i \wedge * \omega_j \,.
\ea
$f^I_i$ and $h_{iI}$ are defined via
\ba
e^{-\frac{K}{2}}\left( \begin{matrix} f^I_i \\ h_{iI} \end{matrix} \right)=\nabla_i \left( \begin{matrix} X^I \\ F_I \end{matrix} \right) = (\partial_i + \partial_i K ) \left( \begin{matrix} X^I \\ F_I \end{matrix} \right) \,.
\ea
Regarding the covariant derivatives, recall that a special K\"ahler manifold $M$ is in particular a Hodge manifold, i.e. comes equipped with a holomorphic hermitian line bundle $L \rightarrow M$ with hermitian connection $\partial_i K$, of which $X^I, F_I$ are local sections \cite{Strominger:1990pd,Ceresole:1995jg,Freed:1997dp}.

The period matrix $\cN$ \cite{deWit:1984pk,Ceresole:1995jg} is specified by the properties
\ba
F_I = \cN_{IJ} X^J  \,\,, \hspace{1cm} h_{Ii} = \bar{\cN}_{IJ}f_i^J \,.
\ea
In terms of a prepotential, it is given by
\ba
\cN_{IJ} &=& \bar{F}_{IJ} + 2i \frac{(\im F)_{IK} X^K (\im F)_{JL}X^L}{X^K (\im F)_{KL} X^L} \,.
\ea
An identity we need is \cite{Ceresole:1995jg}
\ba
f_i^I f_{\bar{\jmath}}^J g_{i \bar{\jmath}} &=& -\frac{1}{2} (\im \cN)^{-1\, IJ} - e^K \bar{X}^I X^J   \,.
\ea

We now evaluate these expressions for our setup. The tree level prepotential describing a single vector multiplet is given by
\ba
F &=& -\frac{1}{3!}C \frac{X_1^3}{X_0}  \label{prepot}\,,
\ea
with $X^0=1, X^1 = b+iv$, where the triple intersection number can be expressed as
\ba
C &=& \int \omega \wedge \omega \wedge \omega \\
&=& \frac{2}{\lambda^{3/2} ||J||} \,.
\ea
It is convenient to express the special geometry quantities in terms of the invariant $C$ and the modulus $v$, in particular,
\ba
||J||^2 &=& \frac{C v^3}{2} \,.
\ea
\ba
e^{-K} &=& \frac{8}{6} \int J \wedge J \wedge J \\
&=& \frac{8}{3} ||J||^2 = \frac{4}{3} C v^3 \,,
\ea
\ba
V &=& \frac{1}{6}Cv^3 \,,
\ea
\ba
g&=& ||\omega||^2 = \frac{1}{v^2} ||J||^2 = \frac{Cv}{2} \,,
\ea
\ba
G &=& \frac{g}{4V} = \frac{3}{4v^2} \,.
\ea
For the cubic prepotential (\ref{prepot}), the period matrix evaluates to
\ba
\cN &=& \left(\begin{array}{cc}
-\frac{1}{6} C \big(2b^3 + i(v^3 + 3 vb^2)\big) & \frac{1}{2} Cb(b+iv) \\ \frac{1}{2} Cb(b+iv) & -\frac{1}{2} C (2b + iv) \end{array}\right) \,.
\ea
The gauge coupling matrix is then
\ba \label{gaugecouplingfromprepot}
(\im \cN)^{-1} &=&
- 8 e^K \left(
\begin{matrix}
1 & b \\
b &  \frac{1}{4G}+b^2 \\
\end{matrix}
\right) \,.
\ea

\section{Conventions and notation} \label{indnot}
\subsection{Indices}
\begin{center}
\begin{tabular}{|l|l|l|l|} \hline
 sector & index & geometric significance & physical significance\\ \hline
\multirow{2}{*}{special K\"ahler} & $i$ & special geometry affine & enumerates vector multiplets \\
& $I$ & special geometry projective & enumerates gauge fields \\ \hline
\multirow{4}{*}{quaternionic K\"ahler} & $u$ & quaternionic coordinate & enumerates matter fields\\ 
& $\alpha$ & $Sp(n)$ holonomy & enumerates hyperinos\\
& $a$ & $Sp(1)$ holonomy & enumerates supersymmetries \\ 
& $x$ &  $\mathfrak{su}(2)$ &\\
& $A$ & local quaternionic & enumerates hypermultiplets \\
& $\Lambda$ & dualized directions & enumerates tensors \\
 \hline
\end{tabular}
\end{center}
\vspace{1cm}
$(\sigma^x)_a{}^b$ are the standard Pauli matrices, i.e. satisify $[\sigma^x, \sigma^y] = 2i \epsilon^{xyz} \sigma^z$. Their indices are raised and lowered by $\epsilon_{ab}$, $\epsilon^{ab}$, with $\epsilon_{ab} \epsilon^{bc}= -\delta_a^c$ and  $\epsilon_{12}=-1$.
$\epsilon_{\alpha \beta}$ denotes the matrix $\epsilon_{ab} \otimes {\rm id}_n$, with ${\rm id}_n$ the $n$ dimensional identity matrix. On tensors (not spinors!) indices are raised and lowered by contraction with $\epsilon_{ab}$ and $\epsilon_{\alpha \beta}$.

\subsection{The Hodge star}
The Hodge star operator is defined, given an orientation $dx^1 \wedge \ldots \wedge dx^n$, via
\ba
*(dx^1 \wedge \ldots \wedge dx^m) &=& dx^{m+1} \wedge \ldots \wedge dx^n \,.
\ea
In particular, on an even dimensional manifold, $(*_2)^2 =1$, $(*_3)^2=-1$, where $*_n$ denotes the Hodge star acting on $n$-forms.
We extend the Hodge star operator linearly to $\bigwedge^n (T^*M)_{\IC}$. With the local expressions $\Omega = dz^1 \wedge dz^2 \wedge dz^3$ and $J = i \sum_i dz^i \wedge d\bar{z}^{\bar{\imath}}$, $dz^i \sim dx^i + i dy^i$, and the standard orientation $dx^1 \wedge dy^1 \wedge \ldots$, the relations $*\Omega = -i \Omega$ and $ *J = \frac{1}{2} J \wedge J$ follow.

\section{Lichnerowicz}  \label{Lichnerowicz}
Consider the equation
\ban
i \frac{\partial g_{a \bar{b}}}{\partial v^i} = \omega_{i\;a\bar{b}} +  v^j \frac{\partial}{\partial v^i}\omega_{j\;a\bar{b}}  \,, \label{metric variation}
\ean
describing the metric variation on a Calabi-Yau manifold under variation of the K\"ahler form. Based on restrictions on metric variations imposed by preserving Ricci flatness, \cite{KashaniPoor:2006si} presents an argument for the vanishing of the second term on the RHS of this equation. This argument must be refined, as it neglects a gauge condition in considering metric variations. We do so here.

The following equation holds for variations of the Ricci tensor (see section 19 of \cite{Lichnerowicz}) under variations $\delta g_{ab} = h_{ab}$ of the metric,
\ban
2 \delta R_{ab} &=& \triangle_L h_{ab} + [\cD(k(h))]_{ab} \,, \label{lichnerowicz}
\ean
where
\ba
k(h)_a &=& \nabla^b h_{ab} - \frac{1}{2} \nabla_a h\,,
\ea 
\ba
(\cD A)_{ab} &=& \nabla_a A_b + \nabla_b A_a \,,
\ea
$h = g^{ab} h_{ab}$, and $\triangle_L$ is the Lichnerowicz Laplacian.
When written out in component form, in terms of covariant derivatives and contractions with the Riemann tensor, $\triangle_L$ acting on $S^n T^*M$ and the ordinary de Rham Laplacian $\triangle =  d^\dagger d + d d^\dagger$ acting on $\bigwedge^n T^*M$ have the same form. By Ebin's slice theorem \cite{Ebin}, we can restrict attention to metric deformations that satisfy $\nabla^b h_{ab} =0$ (this is referred to as de Donder gauge in the physics literature). Upon this gauge choice, \cite{Berger} demonstrates that $h$ is necessarily constant for any variation of an Einstein structure (Lemma 7.1). With this in place, we can conclude from (\ref{lichnerowicz}) that variations of the metric preserving Ricci flatness require 
\ba
\frac{\partial g_{a \bar{b}}}{\partial v^i} dz^a \wedge d\bar{z}^{\bar{b}}
\ea
to be harmonic. As we have not demonstrated that the metric variation (\ref{metric variation}) satisfies the gauge condition $\nabla^b h_{ab} =0$, we are forced to work with the full expression (\ref{lichnerowicz}) for variation of the Ricci form.

To this end, fix a complex structure $J^a_b$. By Yau's theorem, for any K\"ahler class\footnote{In this appendix, we denote the K\"ahler form by $\omega$ to distinguish it clearly from the complex structure $J^a_b$.} $[\omega]$ specified by coordinates $(v^i)$, a unique K\"ahler form $\omega(v)$ exists such that the associated metric
\ba
g_{ab}(v) = -J^c_a \omega_{cb}(v)  
\ea
is Ricci flat. Hence, $\frac{d}{dv} g_{ab}(v)$ must lie in the kernel of the operator appearing in (\ref{lichnerowicz}),
\ba
\triangle_L J^c_a \partial_v \omega_{cb} + \nabla_a \nabla^d J^c_d \partial_v \omega_{cb} + \nabla_b \nabla^d J^c_d \partial_v \omega_{ca} - \nabla_a \nabla_b h&=& 0 \,.
\ea
Passing to complex coordinates, we obtain
\ba
0&=&\triangle_L J^\mu_\rho \partial_v \omega_{\mu \bar{\nu}} + \nabla_\rho \nabla^\sigma J^\mu_\sigma \partial_v \omega_{\mu \bar{\nu}} + \nabla_{\bar{\nu}} \nabla^{\bar{\mu}} J_{\bar{\mu}}^{\bar{\sigma}} \partial_v \omega_{\bar{\sigma} \rho} - \nabla_\rho \nabla_{\bar{\nu}}h \\
&=& i\triangle_L  \partial_v \omega_{\rho \bar{\nu}} + i \nabla_\rho \nabla^\mu \partial_v \omega_{\mu \bar{\nu}} -i \nabla_{\bar{\nu}} \nabla^{\bar{\mu}} \partial_v \omega_{\bar{\mu} \rho} - \nabla_\rho \nabla_{\bar{\nu}}h \\ 
&=& i\triangle  \partial_v \omega_{\rho \bar{\nu}} + i \nabla_\rho \nabla^\mu \partial_v \omega_{\mu \bar{\nu}} +i \nabla_{\bar{\nu}} \nabla^{\bar{\mu}} \partial_v \omega_{\rho \bar{\mu} } - \nabla_\rho \nabla_{\bar{\nu}}h \,.
\ea
We have dropped the ${}_L$ in the last line by the comment above. Anti-symmetrize the last line with regard to $\rho$ and $\bar{\nu}$. The result is 
\ba
0 &=& \triangle \partial_v \omega + (\del \del^\dagger + \bar{\del} \bar{\del}^\dagger) \partial_v \omega \\
 &=& (2 \partial^\dagger \partial + 3 \del \del^\dagger + \bar{\del} \bar{\del}^\dagger) \partial_v \omega \,.
\ea
Now,
\ba
0 &=& ((2 \partial^\dagger \partial + 3 \del \del^\dagger + \bar{\del} \bar{\del}^\dagger) \partial_v \omega , \partial_v \omega) \\
&=& 2 ||\partial \partial_v \omega ||^2 + 3 || \del^\dagger \partial_v \omega||^2 + || \bar{\del}^\dagger \partial_v \omega||^2 \,.
\ea
In particular, $\partial_v \omega$ must be co-closed, hence its Hodge decomposition cannot contain an exact component. As the $\frac{\partial}{\partial v^i}\omega_{j\;a\bar{b}}$ are exact, we conclude
\ba
v^j \frac{\partial}{\partial v^i}\omega_{j\;a\bar{b}} = 0 \,.
\ea

\bibliography{biblio}
\end{document}